\documentclass[final,12pt]{elsarticle}

\usepackage{amssymb}
\usepackage{amsmath}
\usepackage{subcaption}
\usepackage{graphicx}
\usepackage{float}
\usepackage{array}
\usepackage{todonotes}
\usepackage[table]{xcolor}
\renewcommand{\overline}{\bar}

\journal{}

\begin{document}

\begin{frontmatter}

\title{Mathematical modeling on peristaltic flow of a Prandtl fluid with effects of slip conditions and inclined magnetic field}


\author[inst1,inst2]{Sabia Asghar$^{0,}$}

\affiliation[inst1]{organization={Computational Mathematics Group, Department of Mathematics and Statistics, University of Hasselt},
            addressline={Diepenbeek}, 
            country={Belgium}}

\author[inst1,inst2]{Fred J. Vermolen}

\affiliation[inst2]{organization={Data Science Institute (DSI), University of Hasselt},
            addressline={Diepenbeek}, 
            country={Belgium}}
            
\footnotetext{Corresponding author\par
E-mail address: sabia.asghar@uhasselt.be}

\begin{abstract}
The manuscript provides a description of a theoretical analysis of a non-Newtonian Prandtl fluid that is subject to peristaltic flow via an inclined asymmetric channel. {}{We explore the effect of an inclined magnetic field on the peristaltic flow. This is relevant for applications of  fluid flow in narrow, inclined (tilted) tubes that are similar to blood vessels or to the digestive system.} The model also includes thermodynamic aspects such as heat diffusion (the Soret effect) and viscous dissipation as a result of wall-fluid slip conditions,  {}{which help optimize medical devices like lab-on-a-chip systems and dialysis machines.} In this study, the concentration of a generic chemical, temperature and fluid velocity are taken into account through mass, heat and momentum balances, respectively. {}{The solution is approximated by the use of numerical techniques that are suitable for cases with long wavelengths (low frequency) and low Reynold’s numbers. The study also discusses the effects of trapping phenomena, which is a crucial issue from a clinical point of view. The developed insights can be used to improve the understanding of physiological flows in the gastrointestinal tract and in blood vessels. By understanding how the fluid moves and how particles are trapped, these insights help to design better medical pumps and artificial organs. A graphical visualization is provided for the fluid velocity profile, temperature distribution and concentration of a generic chemical. Furthermore, a validation of our numerical results has been provided by means of a comparison with a closed-form solution from a benchmark problem. The Prandtl fluid parameters $\alpha$ and $\beta$ have an opposite impact on the axial velocity. Furthermore, an increase in the Schmidt number, $Sc$, gives a decrease of the concentration of the dissolved chemical. The model predicts that channel inclination has no significant effect on the concentration profile. Furthermore, the model indicates that the Prandtl fluid parameters $\alpha$ and $\beta$ hardly impact the size of the bolus trapped between the streamlines.}
\end{abstract}


\end{frontmatter}


\section{Introduction}
\label{sec1}
Peristalsis is a transport process based on the propelling of a fluid, which is often very viscous, through sinusoidal longitudinal waves in the direction of the channel axis. This mechanism is important in the transport of swallowed food to the stomach, or in the transport of urine from the kidneys to the gall bladder, blood circulation in blood vessels, food digestion and in many more physiological processes. For this reason, the peristaltic mechanism has gained a lot of interest due to its importance in medicine and industry. Furthermore, in industry, one may think of the following devices that work under this principle: fixed bed reactors, tube and finger pumps, heart-lung equipment, open-heart surgery by-pass arteries, and dialysis machines. After the pioneering studies of Latham \cite{1} and Shapiro et al. \cite{2}, several peristaltic transport models of various fluids were considered in a variety of settings. Non-Newtonian fluids exhibit a wide range of flow behaviors, which give rise to a diverse pattern of possible modeling scenarios. Most physiological fluids (blood or chyme) exhibit non-Newtonian behavior. Unfortunately classical Newtonian models fail to capture the real flow dynamics. Furthermore, peristaltic flows in industrial settings comprise pharmaceuticals, cosmetics, and paints. In drug delivery systems, where a substance needs to be released at a controlled rate, the particular flow properties of non-Newtonian flow can be exploited successfully for optimal drug administration. Some studies that treat the peristaltic flow of non-Newtonian fluids have been described in \cite{3}-\cite{7}.

Magneto Hydro Dynamics (MHD) is a mathematical model of flow of an electrically conductive fluid under the influence of a magnetic field \cite{16}. MHD has been studied intensively because of its diverse applications in multiple disciplines in science and industry. Some of these applications are found in magnetic devices for alloy separation, magnetic throttles, DC electric motors, MHD positioning sensors, electro-slags, magnetometers, fuel level indicators and many others. Quantifying the effects of an inclined magnetic field is also helpful in various medical routines that involve, among others, X-rays, MRI, and radiation/laser therapy. Hayat et al. \cite{17} studied the impact of an inclined magnetic field on the peristalsis of a Williamson fluid. The impact of slip features on the peristalsis of a Prandtl nanofluid with an inclined magnetic field was investigated by Hayat et al. \cite{18}. Ramesh \cite{19} considers the effects of heat and mass transfer, as well as the inclination of a magnetic field on the peristaltic flow of a couple stress fluid through a porous medium. This couple stress fluid is a non-Newtonian fluid with a nonsymmetric stress tensor. Furthermore, the effects of heat transfer on the peristaltic motion of an Oldroyd fluid in the presence of an inclined magnetic field have been analyzed by Khan et al. \cite{20}. 

The mixed convection mechanism, observed in multiple technological settings, is mainly caused by mixing effects of natural and forced convection. This mode of convection is prevalent in, among others, heating furnaces, heat exchangers, metallurgical devices, and solar central receivers. Exploring mixed convection in peristalsis also leads to better diagnostic tools for motility disorders and understanding thermal buoyancy effects in small vessels near muscles due to body temperature variations. 
Rahman et al. \cite{8} analyzed the influence of mixed convective heat transfer in the presence of thermal radiation. Hayat et al. \cite{9} studied the impact of mixed convection on the peristaltic transport of nanofluids. In this last-mentioned study, thermal slip and porous media effects are also highlighted. The impact of the addition of gold nanoparticles to the mixed convective Poiseuille flow of a nanofluid through a permeable medium is studied by Aman et al. \cite{10}. Hayat et al. \cite{11} discussed the implications of mixed convection, chemical reactions, and the Hall current in the peristalsis of a Prandtl fluid. 

Although the no-slip boundary condition has been utilized in numerous flow studies, it is not always an appropriate assumption for non-Newtonian fluids. The slip effect occurs at the channel boundaries mainly in the case of rheological fluids. Such fluids include polymers, suspension fluids, lubricating oil, grease, and liquid solutions etc. {}{This slip effect causes a revision of the boundary conditions on the wall. These boundary conditions provide a more accurate description of fluid behavior when traditional no-slip assumptions fail. In particular, slip increases flow rates in tiny channels, reduces adhesion between fluid and wall, and this applies to blood flow in capillaries with endothelial slip effects.} Hayat et al. \cite{12} studied the magnetohydrodynamic (MHD) effects of a hyperbolic tangent model for a nanofluid under slip conditions in an inclined channel.  Ramesh \cite{13} developed analytical solutions with slip conditions on the boundaries for the flow of a Casson fluid. Furthermore, Choudhari et al. \cite{14} studied the peristaltic flow of a Herschel-Bulkley fluid in an elastic tube with slip effects on porous walls. Furthermore, the slip effect for a non-Newtonian fluid has been analyzed by Gudekote et al. \cite{15}.

The Soret effect amounts to the separation between small, light molecules and large, heavy molecules under the influence of a (large) temperature gradient. Typically, this effect is significant when there is a large number of chemical species in the presence of a large temperature gradient. Such circumstances may be encountered in chemical reactors and in cardiovascular diseases. The interaction of peristaltic flow with the Soret effect has been observed in multiple technological, physiological and chemical circumstances. For example, Hayat et al. \cite{21}-\cite{22} investigated the peristaltic transport of nanofluids, including Soret and Dufour effects. Zhao et al. \cite{23} studied MHD in a fractional Maxwell fluid flow with convective heat and mass transfer in line with Soret and Dufour effects.

Hence, interest in the phenomenon of peristaltic flow within the human body and in diverse applications has increased over the past decades. {{} The newly obtained insights in peristaltic flow under diverse circumstances fill a gap in today’s knowledge in this complex (computational) fluid dynamics problem. The current study contains the following innovations: the implementation of the slip boundary condition and the incorporation of an inclined magnetic field to the peristaltic flow. Furthermore, we combine thermodynamic properties such as Joule heating, viscous dissipation, and the Soret separation effect. Incorporating slip boundary conditions for fluid velocity makes the representation of the flow characteristics more realistic. In the current study, we assume that the Reynolds number is low and that the wavelength of the channel walls is large enough to have a relatively tractable model. Numerical approximations are constructed using the finite difference method through the use of the Mathematica software package, and the impact of several model parameters on the flow characteristics is quantified. The novelty of our study is summarized in Table 1. Our study is purely computational, parametric, and conceptual, rather than experimentally or clinically validated. Our study is merely descriptional and not very mathematically rigorous in the sense of formal proofs of the properties that we analyze. Graphical visualizations are given for the flow velocity field, as well as temperature and concentration profiles. Furthermore, we discuss the circumstances under which the trapping phenomenon may occur. The trapping phenomenon is crucially important since it may reflect the case, where food is accumulated in the gullet. The trapping phenomenon is quantified by the concept of a 'bolus', in which chemicals or nutrients are transported uniformly more or less at the wave velocity. This theory may help to understand drug absorption in the intestines.} Important findings, limitations of the study and some future directions are outlined in the last section.

\begin{table}\footnotesize
\centering
\caption{\it {}{Novelty of the study: A comparison of the key features investigated under the current study with those of existing literature}}
\label{Table1}
\begin{tabular}{|>{{}}c|>{{}}c|>{{}}c|>{{}}c|>{{}}c|>{{}}c|>{{}}c|}
\hline
\footnotesize {Key features} & \footnotesize {\citep{45}} & \footnotesize {\citep{11}} & {\citep{25}} & {\citep{41}} & {\citep{46}} &{Current study} \\
\hline
\hline
Prandtl fluid & Yes & Yes & Yes & Yes & Yes & Yes \\
\hline
Inclined magnetic field & No & No & No & No & No & Yes\\
\hline
Slip conditions & No & No & No & No & No & Yes \\
\hline
Mixed convection & No & Yes & No & Yes & No & Yes\\
\hline
Soret effect & No & Yes & No & No & No & Yes \\
\hline
Inclined asymmetric channel & No & No & No & No & No & Yes\\
\hline
Bolus size study & Yes & No & Yes & No & No & Yes\\
\hline
Numerical Solution & Yes & Yes & No & Yes & No & Yes \\
\hline
\end{tabular}
\end{table}

\section{Problem Formulation}
\label{sec2}
{{} The muscles (walls) in the esophagus, stomach, and intestines rhythmically squeeze and contract, and thereby push the partially digested food (a thick, stretchy fluid) forward in the direction of the axis. In this study, we model the thick fluid by a Prandtl fluid (a visco-inelastic non-Newtonian fluid) to represent this food mixture, while the uneven (inclined and asymmetric) channel shape mimics the natural narrowing and widening of the intestines. The inclined magnetic field is considered to control the body's electrical signals that help in muscle contractions and relaxations for fluid motion. Table \ref{Table1} provides a detailed comparison of our study with prior works. By evaluating key features against existing literature, the table clearly demonstrates the distinct contributions offered by the current work.}

\begin{figure}[h!]
\centering
\includegraphics[height=7.5cm]{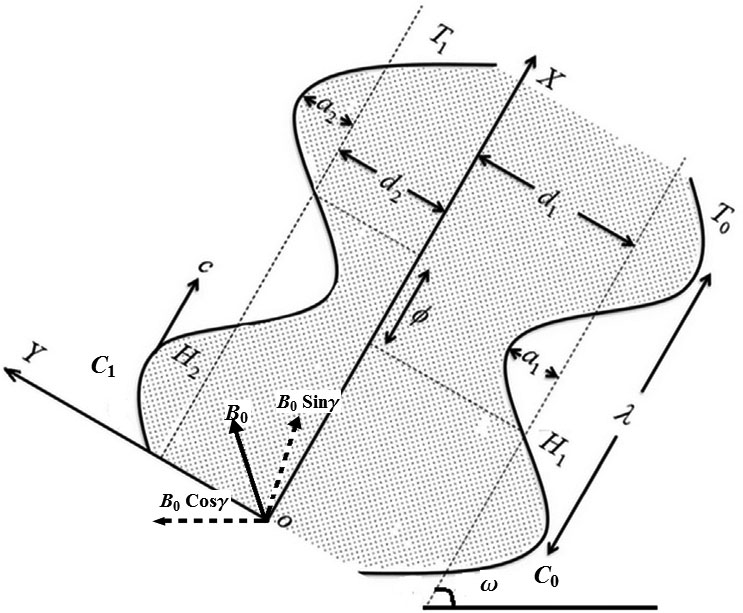}
\caption{\it {}{Schematic depicting the flow geometry (diagram not to scale): the oscillating channel walls in an inclined direction, as well as the magnetic field, which has its own orientation.}}
\label{fig1}
\end{figure}

Therefore, we investigate the peristaltic flow of an incompressible Prandtl fluid within an inclined asymmetric channel of width $d_1+d_2$ (see Fig. \ref{fig1}). Sinusoidal waves propagate along the channel walls with velocity $c$ and wavelength $\lambda$. The direction of wave propagation aligns with the X-axis. The fluid is subject to electrical conduction in the presence of an inclined magnetic field. Moreover, the model incorporates Joule heating, viscous dissipation, and the Soret effect. The geometry of the waves is defined as follows \citep{35}:
\begin{equation}
    Y=H_1(X, \hat{t})=d_1+a_1~cos\left[{\frac{2 \pi}{\lambda}}(X-c \hat{t})\right],
\end{equation}
\begin{equation}
    Y=H_2(X,\hat{t})=-d_2-a_2~cos\left[{\frac{2 \pi}{\lambda}}(X-c \hat{t})+\phi\right],
\end{equation}

in which $a_1$ and $a_2$ are the wave amplitudes (where we assume large wavelengths, that is, $\lambda >>a_1,a_2$), $\hat{t}$ the time and {{}$\phi$ ($0\leq \phi \leq \pi$)} the phase difference. Furthermore, in order to ensure that there is no mechanical impingement between the channel walls, $d_1,d_2,a_1,a_2$ and $\phi$ satisfy
\begin{equation}
    a_1^2+a_2^2+2a_1a_2~cos\phi\leq(d_1+d_2)^2.
\end{equation}
First, the magnetic (Lorentz) force and the expression for Joule heating are formulated. A magnetic field with the following inclination
\begin{equation}
    {\bf{B}}=B_0~sin\gamma ~{\bf{i}}+B_0~cos\gamma~{\bf{j}}+0~{\bf{k}},
    \label{inclinedmagneticfield}
\end{equation}
is applied. Here, $B_0$ and $\gamma$, respectively, represent the strength and angle of inclination of the magnetic field.

The Cauchy stress tensor ${\bf{\tau}}$ for the Prandtl fluid model \cite{26} is given by
\begin{equation}
    {\bf{\tau}}=-P{\bf{I}}+{\bf{S}},
\end{equation}
where $P$ denotes the fluid pressure and the components of $S$ are given by
\begin{equation}
{\bf{S_{ij}}}=\frac{A~\arcsin\left[\sqrt{\sum\limits_{l=1}^{3}\sum\limits_{m=1}^{3}({\frac{e_{lm}e_{ml}}{2N^2}}})\right]}{\sqrt{\sum\limits_{l=1}^{3}\sum\limits_{m=1}^{3}{(\frac{e_{lm}e_{ml}}{2N^2}}})}e_{ij},
\end{equation}
with $e_{ij}$ denoting the components of the strain rate tensor $\bf e$, given by 
\begin{equation}
    {\bf{e}}={\frac{1}{2}}(\nabla {\bf{V}}+(\nabla {\bf{V}})^T),
\end{equation}
where ${\bf V} = [U ~ V]^T$ represents the fluid velocity vector. In addition, ${\bf{S}}$ represents the viscous stress tensor, $\bf I$  the identity  tensor  and $A=\frac{A'}{\mu}$, where $\mu$ denotes the fluid viscosity. Furthermore, $A'$ and $N$ denote the material constants of the Prandtl fluid.

For the electric current density, we use Ohm’s Law, which, in the absence of an electric field \cite{27}, is given by
\begin{equation}
    {\bf J}=k{\bf{(V\times B)}},
\end{equation}
where $k$ is the electrical conductivity and ${\bf{V}}=(U,V,0)$ is the velocity field.
Combining the above expression and equation (\ref{inclinedmagneticfield}) leads to
\begin{align*}
     {\bf{J}}&=k[(U{\bf{i}},V{\bf{j}},0~{\bf{k}})\times (B_0~sin\gamma ~{\bf{i}}+B_0~cos\gamma~{\bf{j}}+0~{\bf{k}}),\\
&=0~{\bf{i}}+0~{\bf{j}}+kB_0(U~cos\gamma-V~sin\gamma){\bf{k}},
\end{align*}
where $\bf i, j,$ and $\bf k$, respectively, represent the unit directions in the coordinates. Furthermore, the Lorentz force $\bf F_L$ per unit volume is expressed by
\begin{align*}
   {\bf{F_L}}={\bf{(J\times B)}}=-(k{B_0}^2 cos\gamma)(U~cos\gamma-V~sin\gamma){\bf{i}}\\+
   (k{B_0}^2 sin\gamma)(U~cos\gamma-V~sin\gamma){\bf{j}}.
\end{align*}
The expression for Joule heating \cite{28} has the following form:
\begin{equation}
    Q_{\text{flow}}=\frac{1}{k} \mathbf{J} \cdot \mathbf{J} = k B_0^2 \left( U^2 \cos^2 \gamma + V^2 \sin^2 \gamma - 2UV \cos \gamma \sin \gamma \right).
\end{equation}
Next, we assess the flow problem where we consider the balance of momentum and incompressibility of the fluid. The continuity equation for an incompressible fluid gives:
\begin{equation}
    \nabla \cdot {\bf{V}}=0,
    \label{E10}
\end{equation}
The balance of momentum is formulated by the adjusted Navier-Stokes equations that incorporate the viscous stress, thermal expansion, magnetic field, and mass expansion {{}\citep{19},\citep{25},\citep{41}}
\begin{align}
\rho \left[ \frac{\partial U}{\partial {\hat{t}}} + U \frac{\partial U}{\partial X} + V \frac{\partial U}{\partial Y} \right] 
& +\frac{\partial P}{\partial X} -( \frac{\partial S_{XX}}{\partial X} + \frac{\partial S_{XY}}{\partial Y} ) =\notag \\
&\quad  \rho g \beta_{t} (\hat{T} - T_0) \sin \omega + \rho g \beta_{c} (\hat{C} - C_0) \sin \omega \notag \\
&\quad - k B_0^2 \cos \gamma \times (U \cos \gamma - V \sin \gamma),
\tag{11}
\end{align}
\begin{align}
\rho \left[ \frac{\partial V}{\partial {\hat{t}}} + U \frac{\partial V}{\partial X} + V \frac{\partial V}{\partial Y} \right] &+ 
\frac{\partial P}{\partial Y} -( \frac{\partial S_{YX}}{\partial X} + \frac{\partial S_{YY}}{\partial Y}) = \notag \\
&\quad \rho g \beta_{t} (\hat{T} - T_0) \cos \omega + \rho g \beta_{c} (\hat{C}- C_0) \cos \omega \notag \\
&\quad + k B_0^2 \sin \gamma \times (U \cos \gamma - V \sin \gamma),
\tag{12}
\label{E12}
\end{align}
where $\rho$ represents the density of mass of the fluid. Furthermore, the first and second terms on the right-hand side of the above equations, respectively, account for thermal and mass expansion. The last term on the right-hand side in these equations accounts for the Lorentz force. Furthermore, $\beta_{t}$ is the thermal expansion coefficient, $\beta_{c}$ is the concentration expansion coefficient, $\hat{T}$ is the fluid temperature, $\omega$ is the channel inclination,  $T_0$ and $T_1$ are the temperatures on the right and left walls, respectively. Let $\hat{C}$ be the concentration of a dissolved chemical in the fluid, and let $C_0$ and $C_1$ be the concentrations of the dissolved chemical in the vicinity of the right and left walls, respectively, and $g$ is the gravity constant. Let $\bf{L}=\nabla~\bf{V}$, then the viscous dissipation of energy is given by the scalar product of $\bf{L}$ and the stress tensor, which is defined by:
\begin{equation}
    {\bf{L:S}}= \sum\limits_{i=1}^{2}\sum\limits_{j=1}^{2} {\bf{L}}_{ij}{\bf{S}}_{ij}.
    \tag{13}
\end{equation}
The balance of energy and mass are given, respectively, by {{}\citep{11},\citep{18}}
\begin{align}
\rho c_{p} \left[ \frac{\partial \hat{T}}{\partial {\hat{t}}} + U \frac{\partial \hat{T}}{\partial X} + V \frac{\partial \hat{T}}{\partial Y} \right] &= 
k^{\ast} \left( \frac{\partial^2 \hat{T}}{\partial X^2} + \frac{\partial^2 \hat{T}}{\partial Y^2} \right) + {\bf{L:S}} \notag \\
&\quad + k B_0^2 \left( U^2 \cos^2 \gamma + V^2 \sin^2 \gamma - 2UV \sin \gamma \cos \gamma \right),
\tag{14}
\label{E14}
\end{align}
where $k^{\ast}$ is the thermal conductivity and $c_p$ the specific heat at constant pressure. The mass balance is given by
\begin{align}
\frac{\partial \hat{C}}{\partial {\hat{t}}} + U \frac{\partial \hat{C}}{\partial X} + V \frac{\partial \hat{C}}{\partial Y} &= 
D \left( \frac{\partial^2 \hat{C}}{\partial X^2} + \frac{\partial^2 \hat{C}}{\partial Y^2} \right) + \frac{D K_{t}}{T_{m}} \left( \frac{\partial^2 \hat{T}}{\partial X^2} + \frac{\partial^2 \hat{T}}{\partial Y^2} \right),
\tag{15}
\label{E15}
\end{align}
in which $D$ is the mass diffusion coefficient and $K_t$ is the thermal diffusion ratio. The boundary conditions in terms of velocity (slip), temperature and concentration on both walls are {{}\citep{38}-\citep{40}}:
\begin{align}
U + \beta_1 (S_{XX}{\bf{n}}_X+S_{XY}{\bf{n}}_Y) &= 0 \quad \text{at} \quad Y = H_1, \notag \\
V - \beta_1 (S_{XY}{\bf{n}}_X+S_{YY}{\bf{n}}_Y) &= 0 \quad \text{at} \quad Y = H_2, \notag \\
\hat{T} + \beta_2 \frac{\partial \hat{T}}{\partial {\bf{n}}} &= T_0 \quad \text{at} \quad Y = H_1, \notag \\
\hat{T} - \beta_2 \frac{\partial \hat{T}}{\partial {\bf{n}}} &= T_1 \quad \text{at} \quad Y = H_2, \notag \\
\hat{C} + \beta_3 \frac{\partial \hat{C}}{\partial {\bf{n}}} &= C_0 \quad \text{at} \quad Y = H_1, \notag \\
\hat{C} - \beta_3 \frac{\partial \hat{C}}{\partial {\bf{n}}} &= C_1 \quad \text{at} \quad Y = H_2.
\tag{16}
\end{align}
where $\beta_1$ is the velocity slip parameter and $\beta_2$, and $\beta_3$
 are the thermal insulation and mass isolation parameters on the wall, respectively. The symmetry of the viscous strain tensor $S$ has been used.
Furthermore, $T_m$ represents the mean temperature. {}{In order to simplify the coupled non-linear equations, we use the Galileo transformation \citep{1}-\citep{2} to convert a complex, unsteady physical problem into a more tractable steady-state problem.} Inherently, the flow is unsteady in the laboratory frame $(X,Y)$ and becomes steady in the wave/moving frame $(x,y)$, which moves with speed $c$ along the waves. These two frames are related by the given transformations:
\begin{align}
U(X,Y,\hat{t}) &= u(X - c\hat{t},y), \quad V(X,Y,\hat{t}) = v(X - c\hat{t},y), \quad u = U - c, \quad v = V, \notag \\
P(X, Y, \hat{t}) &= p(X-c\hat{t},y), \quad \hat{T}(X, Y, \hat{t}) = T(X-c\hat{t},y), \notag \\
 \quad \hat{C}(X, Y, \hat{t}) &= C(X-c\hat{t},y)
~\text{with} \quad x = X - c\hat{t}, \quad y = Y,
\tag{17}
\label{E17}
\end{align}
where $(u,v)$ are the velocity components, $p$, $T$ and $C$ are the pressure, temperature, and concentration, respectively, in the moving frame.

After using the above transformations, Eqs. (\ref{E10})--(\ref{E12}) and Eqs. (\ref{E14})--(\ref{E15}) become
\begin{align}
\frac{\partial u}{\partial x} + \frac{\partial v}{\partial y} &= 0,
\tag{18}
\label{E18}
\end{align}

\begin{align}
\rho \left[ \frac{\partial u}{\partial {\hat{t}}}+(u+c) \frac{\partial u}{\partial x} + v \frac{\partial u}{\partial y} \right] &= 
-\frac{\partial p}{\partial x} + \frac{\partial S_{xx}}{\partial x} + \frac{\partial S_{xy}}{\partial y} 
+ \rho g \beta_{t}(T-T_0) \sin \omega \notag \\
&\quad + \rho g \beta_{c}(C - C_0) \sin \omega \notag \\
&- k B_0^2 \cos \gamma \times \left[ (u+c) \cos \gamma - v \sin \gamma \right],
\tag{19}
\end{align}

\begin{align}
\rho \left[ \frac{\partial v}{\partial {\hat{t}}}+(u+c) \frac{\partial v}{\partial x} + v \frac{\partial v}{\partial y} \right] &= 
-\frac{\partial p}{\partial y} + \frac{\partial S_{yx}}{\partial x} + \frac{\partial S_{yy}}{\partial y} 
+ \rho g \beta_{t}(T-T_0) \cos \omega \notag \\
&\quad + \rho g \beta_{c}(C - C_0) \cos \omega \notag \\
&+ k B_0^2 \sin \gamma \times \left[ (u+c) \cos \gamma - v \sin \gamma \right],
\tag{20}
\end{align}

\begin{align}
\rho c_p \left[ \frac{\partial T}{\partial {\hat{t}}}+(u+c) \frac{\partial T}{\partial x} + v \frac{\partial T}{\partial y} \right] &= 
k^{\ast} \left( \frac{\partial^2 T}{\partial x^2} + \frac{\partial^2 T}{\partial y^2} \right) + {\bf{L:S}} \notag \\
&\quad + k B_0^2 \left[ (u+c)^2 \cos^2 \gamma + v^2 \sin^2 \gamma \right] \notag \\
&\quad - 2 k B_0^2 (u+c) v \cos \gamma \sin \gamma,
\tag{21}
\end{align}

\begin{align}
\left[ \frac{\partial C}{\partial {\hat{t}}}+(u+c) \frac{\partial C}{\partial x} + v \frac{\partial C}{\partial y} \right] &= 
D \left( \frac{\partial^2 C}{\partial x^2} + \frac{\partial^2 C}{\partial y^2} \right) + \frac{D K_T}{T_m} 
\left( \frac{\partial^2 T}{\partial x^2} + \frac{\partial^2 T}{\partial y^2} \right).
\tag{22}
\label{E22}
\end{align}

 Let $\psi$ be the scaled stream function defined for the two-dimensional flow, then
 \begin{equation}
      u=\frac{\partial \psi}{\partial y}, ~v=-\delta \frac{\partial \psi}{\partial x}
      \label{E23}
      \tag{23}
 \end{equation}
 where $\delta=\frac{d_1}{\lambda}$ is the wave number. 
 \section{Non-dimensionalization}
 Scaling gives the following dimensionless variables {{} \citep{41}--\citep{44}
 \begin{align}
\bar{x} &= \frac{x}{\lambda}, \quad \bar{y} = \frac{y}{d_1}, \quad \bar{u} = \frac{u}{c}, \quad \bar{v} = \frac{v}{c}, \quad h_1 = \frac{H_1}{a_1}, \notag \\
h_2 &= \frac{H_2}{d_1}, \quad \bar{t} = \frac{c{\hat{t}}}{\lambda}, \quad \bar{p} = \frac{d_1^2}{\lambda \mu c} p, \quad \delta = \frac{d_1}{\lambda}, \notag \\
d &= \frac{d_2}{d_1}, \quad a = \frac{a_1}{d_1}, \quad b = \frac{a_2}{d_1}, \quad Re = \frac{\rho c d_1}{\mu}, \quad \bar{\beta_{i}} = \frac{\beta_{i}}{d_1} \quad (i = 1, 2, 3), \notag \\
M &= \sqrt{\frac{k}{\mu}} B_0 d_1, \quad \bar{\psi} = \frac{\psi}{c d_1}, \quad Pr = \frac{\mu c_{p}}{k^{\ast}}, \quad {\bf{\bar{S_{ij}}}}= \frac{\mu c}{d_1} {\bf S_{ij}}, \notag \\
\theta &= \frac{T - T_0}{T_1 - T_0}, \quad \sigma = \frac{C - C_0}{C_1 - C_0}, \quad Ec = \frac{c^2}{c_p (T_1 - T_0)}, \notag \\
Gr &= \frac{\rho g d_1^2 \beta_{t} (T_1 - T_0)}{\mu c}, \quad Gc = \frac{\rho g d_1^2 \beta_{c} (C_1 - C_0)}{\mu c}, \quad Sc = \frac{\nu}{D}, \notag \\
Sr &= \frac{\rho D K_T (T_1 - T_0)}{\mu T_{m} (C_1 - C_0)}, \quad Br = \frac{\mu c^2}{k^{\ast} (T_1 - T_0)} = Pr \, Ec
\tag{24}
\end{align}}
In the above expressions, $Re$ is the Reynolds number, $M$ the Hartmann number, $Pr$ is Prandtl number, $\theta$ and $\sigma$ are the dimensionless temperature and concentration, respectively. Furthermore, $Ec$ is the Eckert number, $Gr$ and $Gc$ are the temperature and concentration (modified) Grashoff numbers, respectively, $Sc$ is the Schmidt number, $Sr$ the Soret number, and $Br$ the Brinkman number.

Invoking the above dimensionless parameters, Eqs. (\ref{E18})--(\ref{E22}) after omitting bars, become
\begin{align}
\delta \frac{\partial u}{\partial x} + \frac{\partial v}{\partial y} &= 0,
\tag{25}
\label{E25}
\end{align}

\begin{align}
Re \cdot \delta \left[ (u+1) \frac{\partial u}{\partial x} + \frac{v}{\delta} \frac{\partial u}{\partial y} \right] &= 
-\frac{\partial p}{\partial x} + \delta \frac{\partial S_{xx}}{\partial x} + \frac{\partial S_{xy}}{\partial y} + Gr ~ \theta \sin \omega \notag \\
&\quad + Gc ~ \sigma \sin \omega - M^2 \cos \gamma \left[ (u+1) \cos \gamma - v \sin \gamma \right],
\tag{26}
\end{align}

\begin{align}
Re \cdot \delta^2 \left[ (u+1) \frac{\partial v}{\partial x} + \frac{v}{\delta} \frac{\partial v}{\partial y} \right] &= 
-\frac{\partial p}{\partial y} + \delta^2 \frac{\partial S_{yx}}{\partial x} + \delta \frac{\partial S_{yy}}{\partial y} + \delta ~ Gr ~ \theta \cos \omega \notag \\
&\quad + \delta ~ Gc ~ \sigma \cos \omega + \delta M^2 \sin \gamma \left[ (u+1) \cos \gamma - v \sin \gamma \right],
\tag{27}
\end{align}

\begin{align}
Re \cdot Pr \cdot \delta \left[ (u+1) \frac{\partial \theta}{\partial x} + \frac{v}{\delta} \frac{\partial \theta}{\partial y} \right] &= 
\left( \delta^2 \frac{\partial^2 \theta}{\partial x^2} + \frac{\partial^2 \theta}{\partial y^2} \right) + Br \frac{\partial u}{\partial y} S_{xy} \notag \\
&\quad + M^2 Br \left[ (u+1)^2 \cos^2 \gamma + v^2 \sin^2 \gamma \right] \notag \\
&\quad - 2 M^2 ~ Br (u+1) v \cos \gamma \sin \gamma,
\tag{28}
\end{align}

\begin{align}
Re \cdot \delta \left[ (u+1) \frac{\partial \sigma}{\partial x} + \frac{v}{\delta} \frac{\partial \sigma}{\partial y} \right] &= 
\frac{1}{Sc} \left( \delta^2 \frac{\partial^2 \sigma}{\partial x^2} + \frac{\partial^2 \sigma}{\partial y^2} \right) + Sr \left( \delta^2 \frac{\partial^2 \theta}{\partial x^2} + \frac{\partial^2 \theta}{\partial y^2} \right).
\tag{29}
\end{align} 
With the use of Eq. (\ref{E23}) and the 'long wave length assumption', that is, $\delta \rightarrow 0$, and assuming a low Reynolds number, that is, $Re \rightarrow 0$, one obtains the following approximate relation, upon neglecting higher-order terms in a singular perturbation method:
\begin{align}
\frac{\partial p}{\partial x} &= \frac{\partial S_{xy}}{\partial y} + Gr ~ \theta \sin \omega + Gc ~ \sigma \sin \omega 
- M^2 \cos^2 \gamma \left( \frac{\partial \psi}{\partial y} + 1 \right),
\tag{30}
\label{E30}
\end{align}

\begin{align}
\frac{\partial p}{\partial y} &= 0,
\tag{31}
\label{E31}
\end{align}

\begin{align}
\frac{\partial^2 \theta}{\partial y^2} + Br \frac{\partial^2 \psi}{\partial y^2} S_{xy} + M^2 Br \cos^2 \gamma \left( \frac{\partial \psi}{\partial y} + 1 \right)^2 &= 0,
\tag{32}
\label{E32}
\end{align}

\begin{align}
\frac{1}{Sc} \frac{\partial^2 \sigma}{\partial y^2} + Sr \frac{\partial^2 \theta}{\partial y^2} &= 0.
\tag{33}
\label{E33}
\end{align}
We assume that the quantities are sufficiently smooth and therefore we differentiate Eq. (\ref{E30}) with respect to $y$ and eliminate the pressure gradient term from Eqs. (\ref{E30}) and (\ref{E31}). Then we obtain
\begin{align}
\frac{\partial^2 S_{xy}}{\partial y^2} + Gr \frac{\partial \theta}{\partial y} \sin \omega 
+ Gc \frac{\partial \sigma}{\partial y} \sin \omega - M^2 \cos^2 \gamma \frac{\partial^2 \psi}{\partial y^2} &= 0,
\tag{34}
\end{align}
with
\begin{align}
S_{xy} &= \alpha \frac{\partial^2 \psi}{\partial y^2} + \frac{\beta}{6} \left( \frac{\partial^2 \psi}{\partial y^2} \right)^3,
\tag{35}
\label{E35}
\end{align}

where $\alpha = \frac{A'}{N \mu} \quad \text{and} \quad \beta = \frac{\alpha c^2}{N^2 d_1^2} \quad \text{are the Prandtl fluid parameters}$.

We eventually end up solving Eqs. (\ref{E32})--(\ref{E35}) for $\psi, \theta, \sigma$ and $S_{xy}$. Furthermore, from $\psi$ we recover $u = \frac{\partial \psi}{\partial y}$ ($v = - \delta \frac{\partial \psi}{\partial x}$ is much smaller and neglected).
\par The volumetric flow rate in the laboratory frame is 
\begin{equation}
     Q = \int_{H_1}^{H_2} U(X, Y, t) \, dY,
     \tag{36}
     \label{E36}
\end{equation}
where $H_1$ and $H_2$ are the functions of $X$ and $t$. In the frame that moves with the wave, it becomes 
\begin{equation}
    q = \int_{h_1}^{h_2} u(x, y) \, dy,
    \tag{37}
    \label{E37}
\end{equation}
in which $h_1$ and $h_2$ are functions of $x$ only. From Eqs. (\ref{E17}), (\ref{E36}) and (\ref{E37}), we arrive at the following:
\begin{equation}
   \bar{Q}=q+c~h_1-c~h_2,
    \tag{38}
    \label{E38}
\end{equation}
During period $T$, the average flow in fixed frame $X$ is
\begin{equation}
    Q = {\frac{1}{T}}\int_{0}^{T} \bar{Q} \, dt,
    \tag{39}
    \label{E39}
\end{equation}
After substitution the expression from Eq. (\ref{E38}) and integration of Eq. (\ref{E39}), we get
\begin{equation}
   \bar{Q}=q+c~d_1+c~d_2,.
    \tag{40}
    \label{E40}
\end{equation}
We denote the dimensionless mean flow in the laboratory frame by $F$ and in the moving frame by $\Theta$, so we have
\begin{equation}
    \Theta = \frac{Q}{c d_1}, \quad F = \frac{q}{c d_1} 
    \tag{41}
    \end{equation}
\begin{equation}
        \Theta = F + d + 1 
        \tag{42}
\end{equation}
with
\begin{equation}
 \quad F = \int_{h_2}^{h_1} \frac{\partial \psi}{\partial y} \, dy = \psi(h_1(x)) - \psi(h_2(x))
    \tag{43}
\end{equation}

The non-dimensional form of peristaltic waves in the moving frame becomes
\begin{align}
     h_1(x) &= 1 + a \cos(2\pi x), \notag \\
      h_2(x) &= -d - b \cos(2\pi x + \phi)
     \tag{44}
\end{align}
where  $a,b,d$ and $\phi$ satisfy
\begin{equation}
    \quad a^2 + b^2 + 2ab \cos\phi \leq (1 + d)^2 \tag{45}
\end{equation}
Since $\delta <<1$, we approximate the normal derivatives by $\frac{\partial (.)}{\partial n} \approx \frac{\partial (.)}{\partial y}$, and hence the boundary conditions are given by
\begin{align}
    \psi &= \frac{F}{2}, \quad \frac{\partial \psi}{\partial y} + \beta_1 S_{xy} = -1, \quad \text{on} \, ~y = h_1, \notag \\
\psi &= -\frac{F}{2}, \quad \frac{\partial \psi}{\partial y} - \beta_1 S_{xy} = -1, \quad \text{on} \, ~y = h_2 \tag{46}
\end{align}
\begin{align}
\theta + \beta_2 \frac{\partial \theta}{\partial y} &= 0, \quad \sigma + \beta_3 \frac{\partial \sigma}{\partial y} = 0, \quad \text{on} \, ~y = h_1, \notag \\
\theta - \beta_2 \frac{\partial \theta}{\partial y} &= 1, \quad \sigma - \beta_3 \frac{\partial \sigma}{\partial y} = 1, \quad \text{on} \, ~y = h_2 \tag{47}
\end{align}
The heat transfer coefficient $Z$ on the right wall is given by \cite{32}-\cite{33}:
\begin{equation}
    Z = \left( \frac{\partial h_1}{\partial x} \right) \left( \frac{\partial \theta}{\partial y} \right)_{y \to h_1} \tag{48}
\end{equation}
\section{Results and discussion} \label{Sec3}
\begin{figure}[H]
    \centering
        \includegraphics[width=10.5cm]{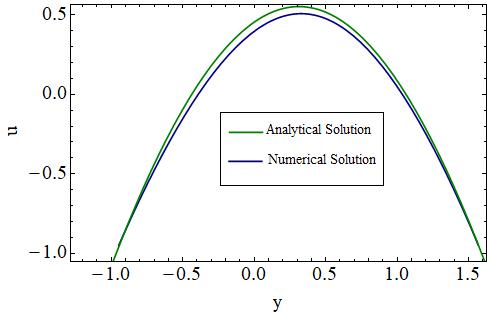}
    \caption{\it {}{Comparison of velocity profile with fixed $x=1, a=0.1, \beta=0.2, b=0.5, d=1, \phi=0.2, \theta=2, \alpha=1.5, M=Sc=\beta_1=\beta_2=\beta_3=\omega=\gamma=Br=Sr=Gr=0 ~\text{and}~ Gc=0$}}
    \label{fig2a}
\end{figure}

In this study, we are interested in the results of the simulations, rather than the numerical solution strategies. The obtained coupled equations are numerically solved after utilizing the NDSolve command in the Mathematica computational software. {}{This command uses the "method of lines" combined with either finite element or finite difference methods. In this study, we use the finite difference method to solve the boundary value problem. The magnitudes of the parametric values/ranges have been taken from the literature, where they have already been used in previous simulations. The values can be found in Table \ref{Table2} along with their references. Furthermore, for the sake of consistency the numerical solution for the velocity profile obtained in this study is also compared with the semi-analytical solution obtained by Akbar et al. \citep{45}. Fig.
\ref{fig2a} shows the comparison of the numerical solution with that of the analytical solution in \citep{45} with the fixed parametric values. We observe that both analytical and numerical solutions show good agreement. The numerical solution captures the overall trend and is slightly smoother, which guarantees the reliability of the solution.} Subsequently, numerical approximations are graphically illustrated for velocity, temperature, concentration, heat transfer rate, and streamline patterns to represent the solutions of the coupled system of equations under the influence of various parameters.
\begin{table}
\centering
\caption{\it {}{Ranges of involved parameters chosen from the existing literature}}
\label{Table2}
\begin{tabular}{|>{{}}c|>{{}}c|>{{}}c|}
\hline
Parameter & Description & Source \\
\hline
\hline
$M$ & Hartmann number & \citep{11}, \citep{25} \\ 
\hline
$Gr$ & Grashoff number & \citep{9} \\ 
\hline
$Gc$ & Modified Grashoff number & \citep{47} \\ 
\hline
$Sc$ & Schmidt number &  \citep{18}\\
\hline
$Sr$ & Soret number &  \citep{22}\\
\hline
$Br$ & Brinkman number &  \citep{11}, \citep{25}\\
\hline
$\alpha$ and $\beta$ & Prandtl fluid parameters &  \citep{11}, \citep{25}\\
\hline
$\beta_1$ & Velocity slip parameter &  \citep{9}\\
\hline
$\beta_2$ & Temperature slip parameter &  \citep{9}\\
\hline
$\beta_3$ & Concentration slip parameter &  \citep{15}\\
\hline
$\gamma$ & Magnetic field inclination &  \citep{17}\\
\hline
$\omega$ & Channel inclination &  \citep{33}\\
\hline
\end{tabular}
\end{table}

\subsection{Impact on the velocity profile}
\begin{figure}[H]
    \centering
    \begin{subfigure}{0.49\textwidth}
        \centering
        \includegraphics[width=\linewidth]{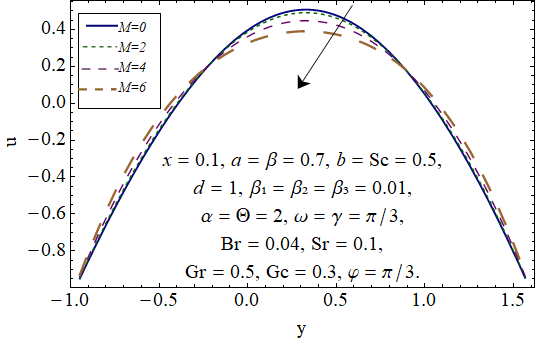}
        \caption{\textit{Velocity for different values of $M$}}
        \label{fig2}
    \end{subfigure}
    \hfill
    \begin{subfigure}{0.49\textwidth}
        \centering
        \includegraphics[width=\linewidth]{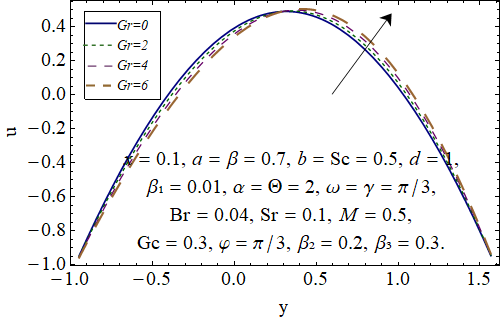}
        \caption{\it Velocity for different values of $Gr$}
        \label{fig3}
    \end{subfigure}
     \hfill
    \begin{subfigure}{0.49\textwidth}
        \centering
        \includegraphics[width=\linewidth]{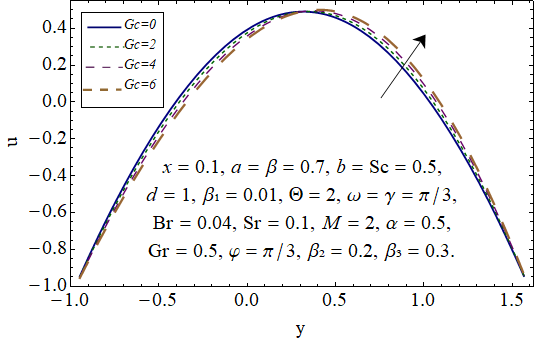}
        \caption{\it Velocity for different values of $Gc$}
        \label{fig4}
    \end{subfigure}
     \hfill
    \begin{subfigure}{0.49\textwidth}
        \centering
        \includegraphics[width=\linewidth]{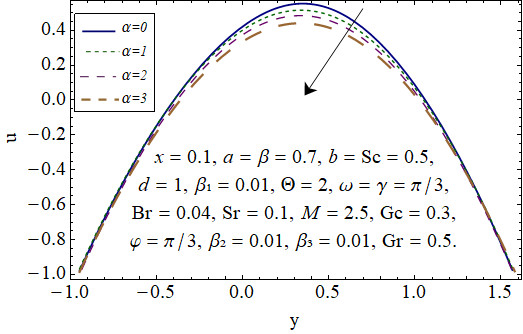}
        \caption{\it Velocity for different values of $\alpha$}
        \label{fig5}
    \end{subfigure}
    \hfill
    \begin{subfigure}{0.49\textwidth}
        \centering
        \includegraphics[width=\linewidth]{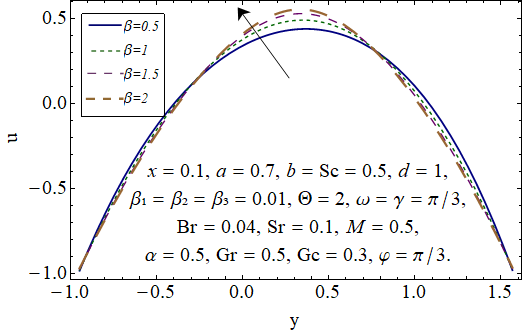}
        \caption{\it Velocity for different values of $\beta$}
        \label{fig6}
    \end{subfigure}
    \hfill
    \begin{subfigure}{0.49\textwidth}
        \centering
        \includegraphics[width=\linewidth]{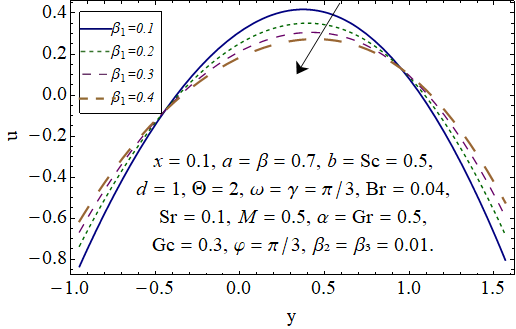}
        \caption{\it Velocity for different values of $\beta_1$}
        \label{fig7}
    \end{subfigure}
    \hfill
    \begin{subfigure}{0.49\textwidth}
        \centering
        \includegraphics[width=\linewidth]{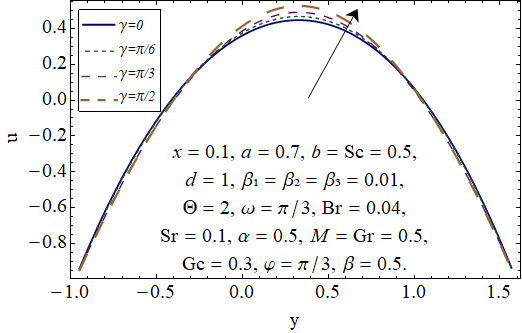}
        \caption{\it Velocity for different values of $\gamma$}
        \label{fig8}
    \end{subfigure}
    \hfill
    \begin{subfigure}{0.49\textwidth}
        \centering
        \includegraphics[width=\linewidth]{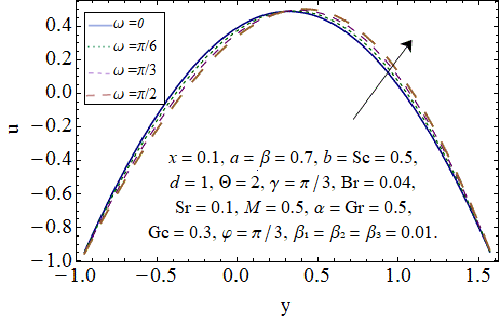}
        \caption{\it Velocity for different values of $\omega$}
        \label{fig9}
    \end{subfigure}
    \caption{Influence of various physical parameters on the velocity profile.}
\end{figure}

Fig. (\ref{fig2}) shows the impact of the Hartmann number $M$ (which signifies the ratio of electromagnetic force and viscous force) on the (horizontal) velocity. It can be seen that, for increasing values of the Hartmann number $M$, there is a decrease of the fluid velocity near the center-line of the channel. This trend results from the fact that for a larger strength of magnetic field $B_0$, the Lorentz force amplifies, which acts as a retardation force, consequently leading to a decrease in fluid velocity. Hence an increase in $M$ declines fluid motion which can be seen in Fig. (\ref{fig2}). It is observed in Fig. (\ref{fig3}) that for increasing values of the Grashoff number $Gr$, the fluid velocity is asymmetric and larger near the upper peristaltic wall. This effect is due to the reduction in drag force, whereas the fluid velocity is smaller towards the lower wall. Since the Grashoff number $Gr$ represents the ratio of buoyancy forces and viscous forces, it is a way to quantify the opposing forces. Similar behavior is observed for the modified Grashoff number $Gc$ (see Fig. (\ref{fig4})). Figs. (\ref{fig5}) and (\ref{fig6}) illustrate the quantitative impact of the parameters of the Prandtl fluid on the velocity distribution. The velocity of the fluid decays at the center of the channel when the fluid parameter $\alpha$ increases. The previous work {{} \cite{11}} validates our result for the fluid parameter $\alpha$. Furthermore, the velocity of the fluid increases when the Prandtl fluid parameter $\beta$ increases. Higher values of the slip parameter $\beta_1$ result in a reduction of the velocity of the fluid near the center of the channel. Since velocity slip provides a deviation between the wall's wave velocity and the fluid velocity adjacent to the wall, a reduction in the fluid velocity is observed. Note that $\beta_1 = 0$ models a non-slip condition. This effect of a less steep velocity profile can be seen in Fig. (\ref{fig7}). Furthermore, it is important to observe that if $\beta_1 \rightarrow 0$ then $u \rightarrow -1$, that is, $U \rightarrow c$. Figs. (\ref{fig8}) and (\ref{fig9}) indicate the impact of the inclination angle of the magnetic field $\gamma$ and the channel inclination angle $\omega$, respectively. The angle inclination of the magnetic field enlarges the velocity near the center (see Fig. (\ref{fig8})). As a result of gravity, the channel inclination angle $\omega$ increases the velocity towards the upper wall and slows the fluid flow near the lower wall. This results in an asymmetric velocity profile (see Fig. (\ref{fig9})).

\subsection{Impact on the temperature profile}
\begin{figure}[H]
    \centering
    \begin{subfigure}{0.49\textwidth}
        \centering
        \includegraphics[width=\linewidth]{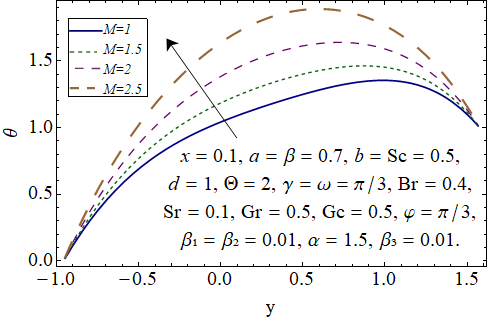}
        \caption{\textit{Temperature for different values of $M$}}
        \label{fig10}
    \end{subfigure}
    \hfill
    \begin{subfigure}{0.49\textwidth}
        \centering
        \includegraphics[width=\linewidth]{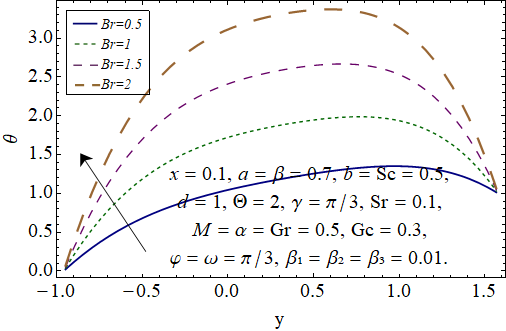}
        \caption{\it Temperature for different values of $Br$}
        \label{fig11}
    \end{subfigure}
     \hfill
    \begin{subfigure}{0.49\textwidth}
        \centering
        \includegraphics[width=\linewidth]{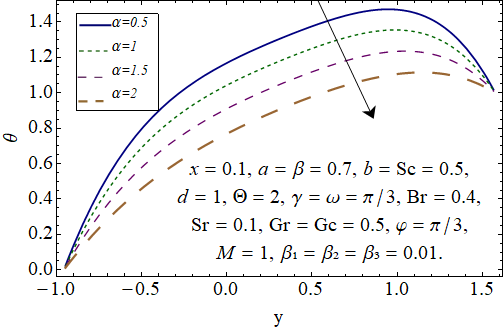}
        \caption{\it Temperature for different values of $\alpha$}
        \label{fig12}
    \end{subfigure}
     \hfill
    \begin{subfigure}{0.49\textwidth}
        \centering
        \includegraphics[width=\linewidth]{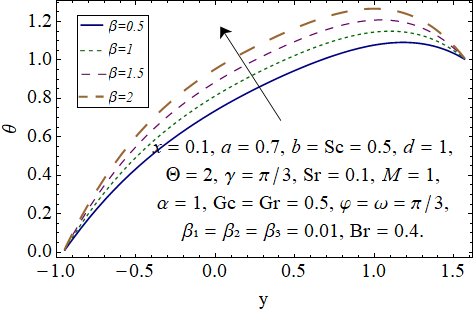}
        \caption{\it Temperature for different values of $\beta$}
        \label{fig13}
    \end{subfigure}
    \hfill
    \begin{subfigure}{0.49\textwidth}
        \centering
        \includegraphics[width=\linewidth]{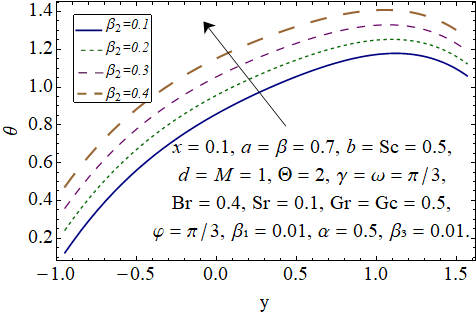}
        \caption{\it Temperature for different values of $\beta_2$}
        \label{fig14}
    \end{subfigure}
    \hfill
    \begin{subfigure}{0.49\textwidth}
        \centering
        \includegraphics[width=\linewidth]{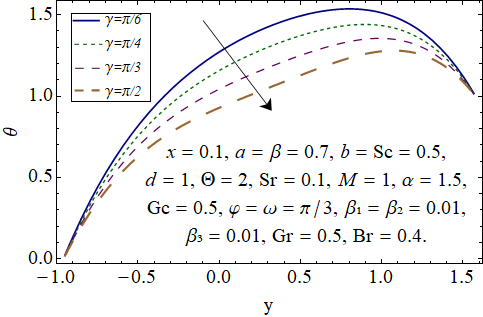}
        \caption{\it Temperature for different values of $\gamma$}
        \label{fig15}
    \end{subfigure}
    \caption{Influence of various physical parameters on temperature profile.}
\end{figure}

The effects of various parameters on the temperature of the fluid are plotted in Figs. (\ref{fig10})--(\ref{fig15}). The temperature exhibits a significant increasing trend with increasing Hartmann number $M$. This is caused by the Joule heating effect, since this effect generates more heat when the strength of the magnetic field increases, and hence the temperature increases, as shown in Fig. (\ref{fig10}). Fig. (\ref{fig11}) shows the impact of increasing the Brinkman number $Br$ on temperature. A larger Brinkman number corresponds to a greater internal resistance between fluid molecules primarily due to viscous dissipation. Therefore, increasing the Brinkman number causes a sharper temperature increase. The validity of this numerical result is supported by the studies by{{}\cite{11} and \cite{12}}. Fig. (\ref{fig12}) illustrates that when the Prandtl fluid parameter $\alpha$ increases, there is a decrease in temperature, while the opposite effect is observed towards the second fluid parameter $\beta$ (as shown in Fig. (\ref{fig13})). This is a consequence of the reduction in the dynamic viscosity of the fluid. Furthermore, there is a direct relationship between the fluid velocity and temperature, which leads to the similar nature of the observations.  Upon increasing the temperature insulation parameter $\beta_2$ (see the boundary condition), the (dimensionless) temperature $\theta$ increases. This follows from the fact that the kinetic energy of the fluid molecules increases as a result of slipperiness, leading to an increase in the (dimensionless) temperature $\theta$ (see Fig. (\ref{fig14})). Fig. (\ref{fig15}) shows the impact of the angle of inclination of the magnetic field on the temperature profile. As the angle increases, it leads to the reduction in drag force which enhances the net flow and reduces the temperature. The relevance of this phenomenon is confirmed {{}in \cite{29}}.

\subsection{Impact on the concentration profile}
\begin{figure}[H]
    \centering
    \begin{subfigure}{0.49\textwidth}
        \centering
        \includegraphics[width=\linewidth]{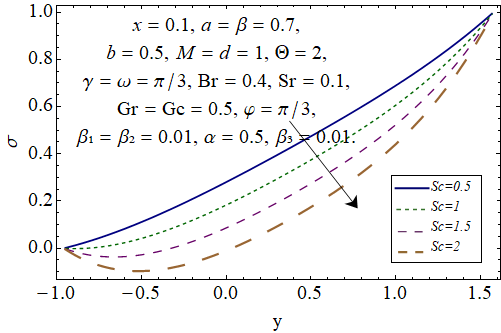}
        \caption{\textit{Concentration for different values of $Sc$}}
        \label{fig16}
    \end{subfigure}
    \hfill
    \begin{subfigure}{0.49\textwidth}
        \centering
        \includegraphics[width=\linewidth]{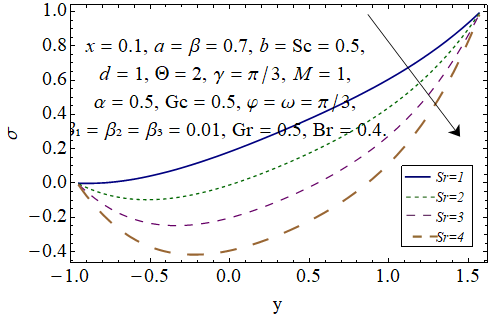}
        \caption{\it Concentration for different values of $Sr$}
        \label{fig17}
    \end{subfigure}
     \hfill
    \begin{subfigure}{0.49\textwidth}
        \centering
        \includegraphics[width=\linewidth]{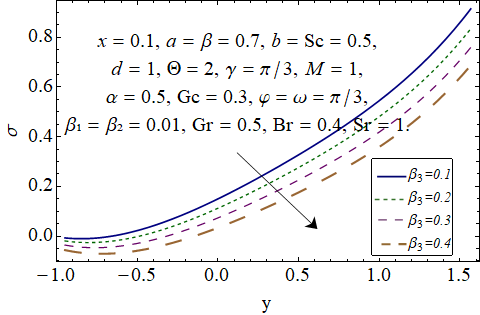}
        \caption{\it Concentration for different values of $\beta_3$}
        \label{fig18}
    \end{subfigure}
     \hfill
    \begin{subfigure}{0.49\textwidth}
        \centering
        \includegraphics[width=\linewidth]{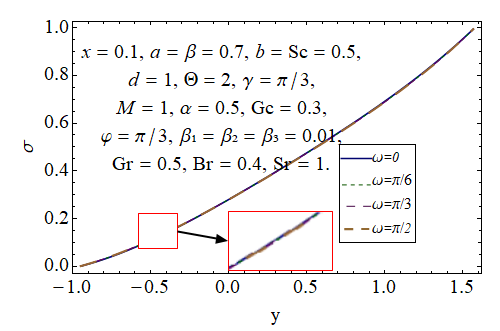}
        \caption{\it Concentration for different values of $\omega$}
        \label{fig19}
    \end{subfigure}
    \caption{Influence of various physical parameters on concentration profile.}
\end{figure}

In this subsection, the impact of the Schmidt number $Sc$, the Soret number $Sr$ and the concentration slip parameter $\beta_3$ on the concentration profile is investigated. The Schmidt number $Sc$, describes the ratio between the rates of viscous (momentum) diffusion and molecular (mass) diffusion. For greater values of $Sc$, the relative proportion of the mass diffusion rate decreases, which therefore forces the particles to move away. Therefore, this causes the concentration to reduce, as shown in Fig. (\ref{fig16}). A decline in the concentration profile is observed with increasing Soret number $Sr$, see Fig. (\ref{fig17}). The Soret number provides a measure of how the concentration of dissolved particles is affected by changes of the temperature gradient, hence an increase causes a reduction in the concentration field in the case of large deviations of temperature gradients (that is, diffusive heat flows). Fig. (\ref{fig18}) shows the impact of the mass isolation parameter $\beta_3$ on the concentration. A decrease in concentration profile is observed for increasing values of $\beta_3$. The increase in $\beta_3$ reflects the lower resistive impact due to the channel walls on fluid particles due to slipperiness. Hence, it causes a reduction in the mass transfer rate. Fig. (\ref{fig19}) (see the zoomed-in part) shows that the model predicts that there is no effect of the inclination of the channel $\omega$ on the concentration profile.

\subsection{Impact on the heat transfer coefficient}
\begin{figure}[H]
    \centering
    \begin{subfigure}{0.49\textwidth}
        \centering
        \includegraphics[width=\linewidth]{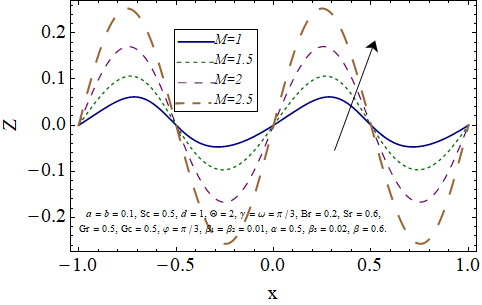}
        \caption{\textit{$Z$ for different values of $M$}}
        \label{fig20}
    \end{subfigure}
    \hfill
    \begin{subfigure}{0.49\textwidth}
        \centering
        \includegraphics[width=\linewidth]{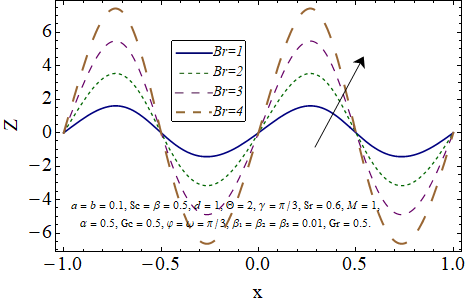}
        \caption{\it $Z$ for different values of $Br$}
        \label{fig21}
    \end{subfigure}
     \hfill
    \begin{subfigure}{0.49\textwidth}
        \centering
        \includegraphics[width=\linewidth]{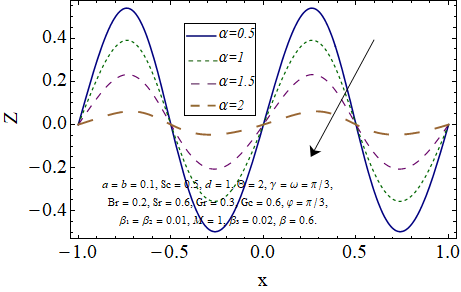}
        \caption{\it $Z$ for different values of $\alpha$}
        \label{fig22}
    \end{subfigure}
     \hfill
    \begin{subfigure}{0.49\textwidth}
        \centering
        \includegraphics[width=\linewidth]{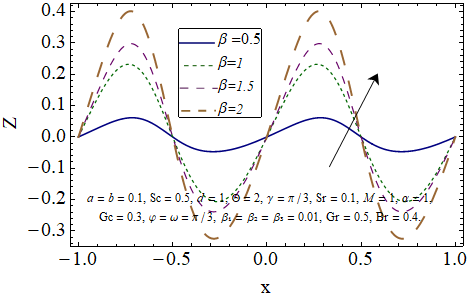}
        \caption{\it $Z$ for different values of $\beta$}
        \label{fig23}
    \end{subfigure}
    \hfill
    \begin{subfigure}{0.49\textwidth}
        \centering
        \includegraphics[width=\linewidth]{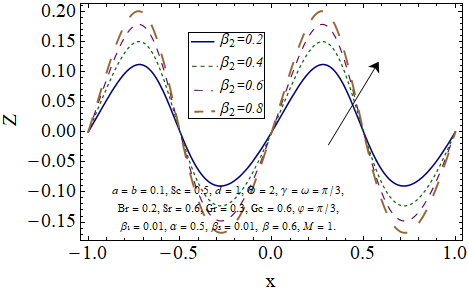}
        \caption{\it $Z$ for different values of $\beta_2$}
        \label{fig24}
    \end{subfigure}
    \hfill
    \begin{subfigure}{0.49\textwidth}
        \centering
        \includegraphics[width=\linewidth]{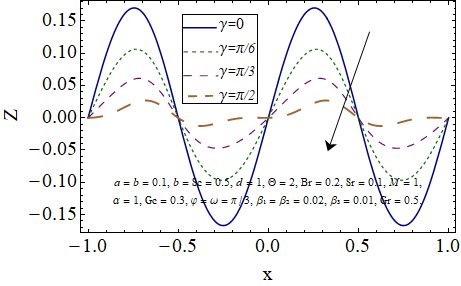}
        \caption{\it $Z$ for different values of $\gamma$}
        \label{fig25}
    \end{subfigure}
    \caption{Influence of various physical parameters on heat transfer coefficient.}
\end{figure}

Figs. (\ref{fig20})--(\ref{fig25}) visualize the impact of various parameters on the heat transfer rate $Z$. The graphs show that due to the contraction and relaxation of the channel walls, the heat transfer coefficient $Z$ exhibits oscillatory behavior in the $x-$direction. Furthermore, the most significant effects are observed towards the center of the channel. Observations from Fig. (\ref{fig20}) reveal that ascending values of the Hartmann number $M$, enhances the heat transfer rate. This is a result of a reduction in fluid viscosity and a gain in kinetic energy, which speeds up the less viscous fluid particles. This consequently leads to an increase in the heat transfer rate (see Fig. (\ref{fig20})). A similar impact is observed for larger Brinkman numbers $Br$ (see Fig. (\ref{fig21})). Variations in the parameters of the Prandtl fluid $\alpha$ and $\beta$ show reverse behavior (see Figs. (\ref{fig22}) and (\ref{fig23})). The results obtained for the fluid parameters are similar to those discussed {{}in Alsaedi \cite{25}}, which validate the current results.
Fig. (\ref{fig24}) shows the impact of the thermal insulation parameter $\beta_2$ on $Z$.  An increase in $\beta_2$ amplifies the rate of heat transfer. Increasing the angle of inclination of the magnetic field $\gamma$ reduces  the  magnitude  of the heat transfer coefficient $Z$ (see Fig. (\ref{fig25})).

\subsection{Trapping}
In peristaltic flow where the sinusoidal walls move at a predefined velocity, there may be regions around the thickest parts of the channel where the fluid flows in a loop (closed contour in an axial-symmetric setting (which gives a toriodal shape in 3D) or in a rectangular channel setting). Streamlines represent contour lines of the stream function $\phi$, that is, lines where $\phi$ is constant. Since the difference between the values of the stream function at two different points defines the amount of flow through any curve connecting these two points, it follows that the streamlines are parallel to the fluid velocity. Hence, closed contours of the stream function represent loops with no outflow. However, towards the central axis of the channel, the streamlines are no longer closed, and hence the fluid is transported through the tube at a smaller speed. This difference in speed is a consequence of the viscous drag. The open, non-empty, simply connected domain of computation is given by $\Omega \subset \mathbf{R}^2$. Let $\mathcal{C}_0^c$ be a closed streamline where $\psi = c$ encloses point $(x_0,y_0) \in \Omega$ then
\begin{align*}
\mathcal{C}_0^c := \{(x,y) \in \Omega:~\psi = \text{$c$, }  \text{ is closed and encloses }(x_0,y_0) \text{ for } c \in \mathbb{R} \},
\end{align*}
then $\mathcal{C}_0^c$ is a closed subset in $\Omega$ because its complement is open. Let $\overline{\mathcal{C}}_0^c$ be the union of $\mathcal{C}_0^c$ and its portion of $\Omega$ it encloses; then $\overline{\mathcal{C}}_0^c$ is a closed subset in $\Omega$. Note that if $\overline{\mathcal{C}}_0^c = \emptyset$ or $\overline{\mathcal{C}}_0^c = \Omega$, then $\overline{\mathcal{C}}_0^c$ is also an open subset (hence, clopen) in $\Omega$. Hence, the union of all $\overline{\mathcal{C}}_0^c$ is closed, that is, $\displaystyle{\cup_c \overline{\mathcal{C}}_0^c}$ is closed. Let $|\mathcal{C}_0^c|$ denote the area that is enclosed by $\mathcal{C}_0^c$, then the maximum of the area enclosed by a streamline is defined by
\begin{align*}
|\hat{\mathcal{C}}_0| = \max \{~ |\mathcal{C}_0| ~ : \mathcal{C}_0 \text{ is closed, and encloses } (x_0,y_0)\}.
\end{align*}
This maximum exists since $\cup_c \overline{\mathcal{C}}_0^c$ is closed.
Then the {\em bolus} around the point $(x_0,y_0)$ is uniquely defined by
\begin{align*}
\hat{\mathcal{C}}_0 = \{\mathcal{C}_0~:~|\mathcal{C}_0| = |\hat{\mathcal{C}}_0|\}.
\end{align*}
Uniqueness is warranted since if there were two boluses around $(x_0,y_0)$, then the boluses would intersect since they enclose the same area and both enclose $(x_0,y_0)$. This would imply that the streamlines would intersect, and hence the flow velocity would have multiple directions, which is impossible.
If there is no bolus, that is, there is no closed streamline around $(x_0,y_0)$, then $\mathcal{C}_0 = \emptyset$, and hence $|\hat{\mathcal{C}}_0| = 0$. The definition of bolus is in line with the definition of Ehsan et al. \cite{34}, though with minor adjustment. The existence of a bolus is critical and physically corresponds to the fluid particles that are trapped in a region of maximal thickness of the channel. This phenomenon is commonly referred to as 'trapping'.

\begin{figure}[H]
    \centering
    \begin{subfigure}{0.49\textwidth}
        \centering
        \includegraphics[width=\linewidth]{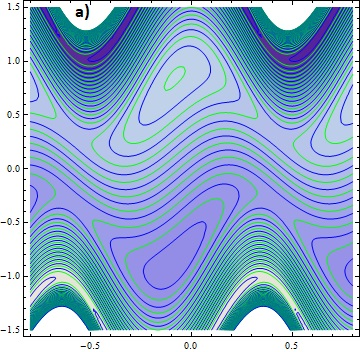}
        \caption{\it Streamlines pattern for fluid parameter $\alpha$=5}
        \label{fig26a}
    \end{subfigure}
     \hfill
    \begin{subfigure}{0.49\textwidth}
        \centering
        \includegraphics[width=\linewidth]{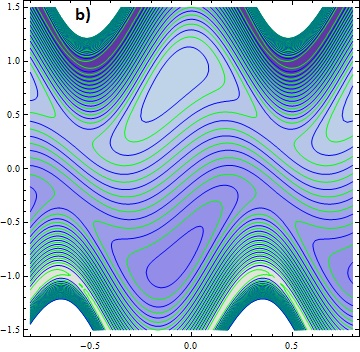}
        \caption{\it Streamlines pattern for fluid parameter $\alpha$=6}
        \label{fig26b}
    \end{subfigure}
    \caption{\it Streamlines profile with $a=\beta=0.7, b=Sc=Gr=0.5, d=1, \theta=2, \omega=\gamma=\pi /3, Br=0.04, Sr=0.1, M=2.5, Gc=0.3, \phi=\pi /3, ~\text{and}~ \beta_1=\beta_2=\beta_3=0.01.$}
    \label{fig26}
\end{figure}

\begin{figure}[H]
    \centering
    \begin{subfigure}{0.49\textwidth}
        \centering
        \includegraphics[width=\linewidth]{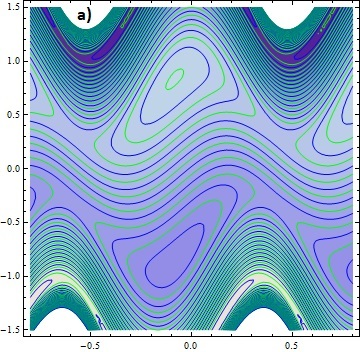}
        \caption{\it Streamlines pattern for fluid parameter $\beta$=1}
        \label{fig27a}
    \end{subfigure}
     \hfill
    \begin{subfigure}{0.49\textwidth}
        \centering
        \includegraphics[width=\linewidth]{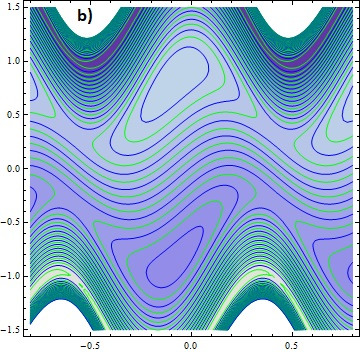}
        \caption{\it Streamlines pattern for fluid parameter $\beta$=2}
        \label{fig27b}
    \end{subfigure}
    \caption{\it Streamlines profile with $a=0.7, d=1, \theta=2, \omega=\gamma=\pi /3, Br=0.04, Sr=0.1, M=b=0.5, \alpha=Sc=Gr=0.5, Gc=0.3, \phi=\pi /3, ~\text{and}~ \beta_1=\beta_2=\beta_3=0.01.$}
    \label{fig27}
\end{figure}

\begin{figure}[H]
    \centering
    \begin{subfigure}{0.49\textwidth}
        \centering
        \includegraphics[width=\linewidth]{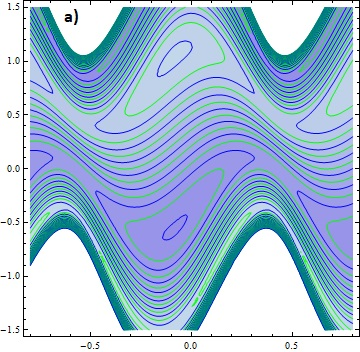}
        \caption{\it Streamlines pattern for velocity slip parameter $\beta_1$=1}
        \label{fig28a}
    \end{subfigure}
     \hfill
    \begin{subfigure}{0.49\textwidth}
        \centering
        \includegraphics[width=\linewidth]{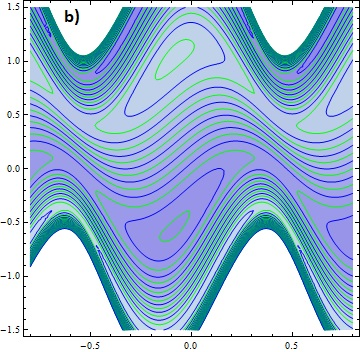}
        \caption{\it Streamlines pattern for velocity slip parameter $\beta_1$=2}
        \label{fig28b}
    \end{subfigure}
    \caption{\it Streamlines profile with $a=\beta=0.7, b=Sc=0.5, d=1, \beta_2=\beta_3=0.01, \theta=2, \omega=\gamma=\pi /3, Br=0.04, Sr=0.1, M=\alpha=0.5, Gr=0.5, \phi=\pi /3 ~\text{and}~ Gc=0.3$}
    \label{fig28}
\end{figure}

Hence in trapping, we observe a bolus that moves at the same wave speed as the channel speed, which is pushed ahead due to contraction and expansion of progressive wave. Figs. (\ref{fig26})--(\ref{fig28}) illustrate the streamline patterns for various values of some physical parameters under fixed values for the other parameters. Figs. \ref{fig26}(a)--(b) and Figs. \ref{fig27}(a)--(b) illustrate the streamlines for different values of the Prandtl fluid parameters $\alpha$ and $\beta$. It is observed that the area of the (trapped) bolus hardly reduces with increasing values of fluid parameters $\alpha$ and $\beta$. The current results are similar to those obtained {{}by Akbar \cite{30}}. The increase in the values of the velocity slip parameter reduces the size of the bolus (see Fig. \ref{fig28}(a)--(b)). This fact indicates that the bolus will be larger in a non-slip scenario {{}\cite{31}}.

\section{Discussion and conclusions}
{{}We conclude the study of peristaltic fluid flow for an inclined MHD Prandtl fluid under the assumption of long wall wavelength $\lambda$ ($\delta \rightarrow 0$), in which Joule heating, viscous dissipation and slip conditions are incorporated. Analysis improves our understanding of peristalsis in physiological processes, such as heart and lung function, blood circulation, digestion, and dialysis. The key findings of this model provide valuable insight into the dynamics of biofluid in both biological and industrial applications. The main results are summarized as follows.}
\begin{itemize}
    \item The Prandtl fluid parameters $\alpha$ and $\beta$ have contrasting behavior i.e. decreasing and increasing, respectively, towards the axial velocity near the center of the asymmetric channel.
    \item  The contrasting dependence of the global temperature field on both the fluid parameters, $\alpha$ and $\beta$, is examined. This dependency modifies the energy distribution throughout the channel.
    \item The velocity slip parameter $\beta_1$ releases the fluid motion toward the center and hence increases the fluid's mobility and affects the core flow dynamics.
    \item  The inclination of the magnetic field makes the temperature diminish, whereas the magnitude of the fluid velocity increases.
    \item The magnitude of the velocity decreases with increasing Hartmann number $M$, while the temperature increases as a result of the Joule heating effect.
    \item  Increasing the thermal boundary insulation parameter $\beta_2$, increases the fluid temperature and the magnitude of the heat transfer rate $Z$. This increase further enhances thermal diffusion on the channel boundaries.
    \item The concentration of the chemical decreases with increasing Schmidt number $Sc$. 
    \item It is observed that the parameters of the Prandtl fluid $\alpha$ and $\beta$ hardly reduce the volume (area) of the bolus trapped between the streamlines.
\end{itemize}

\subsubsection*{{}Limitations of the model}
{{}We note that this study is based on the assumptions of large wall wavelength, small Reynolds number, and constant wall velocity. Furthermore, the current study has been done for a planar channel. These assumptions allowed to cast the set of partial differential equations into a tractable form so that the solution could be approximated easily by a finite difference procedure in Mathematica. In order to make the model more generic, one has to assess the full incompressible Navier-Stokes system, as well as the full conservation laws for energy and mass, and introduce axial symmetry for 'straight' wavelike channels, or even rely on a full 3D solution if channels are curved.}
\subsubsection*{{}Future directions}
The next extension would be to introduce rotational symmetry so that the arteries of tube-like transport systems in the human body or peristaltic pumps can be simulated. Akbar \cite{30} modelled blood flow in a peristaltic artery. This study sparked our interest to explore this further. {{} Another extension would be the incorporation of electrically charged particles in the fluid, which by their convective transport, generate a magnetic field (according to Biot-Savart's Law), which amounts to magnetic induction. This induction generates another magnetic field, which can be modeled by a conservation equation that involves induction and magnetic diffusion. Magnetic diffusivity, $\eta$, represents the denominator in the magnetic Reynolds number ($Rm = \frac{U d}{\eta}$), determines the ratio between magnetic induction (being proportional to fluid motion) and magnetic diffusion. This effect has not yet been incorporated in the current study.}

\section*{Acknowledgments}
We are grateful for the financial support from the Higher Education Commission (HEC) of Pakistan in the framework of project: 1(2)/HRD/OSS-III/BATCH-3/2022/HEC/527.

\section*{Conflict of interest}
The authors declare that they have no known competing financial interests or personal relationships that could
have appeared to influence the work reported in this paper.


\begin{thebibliography}{}

\bibitem{1}
Latham TW (1966). Fluid motion in peristaltic pump. MS Thesis, MIT Cambridge, MA.

\bibitem{2}
Shapiro AH, Jaffrin MY, Weinberg SL (1969). Peristaltic pumping with long wavelengths at low Reynolds number. Journal of Fluid Mechanics 37 (4):799--825.

\bibitem{3}
Taghvaei M, Amani E (2023). Wall-modeled large-eddy simulation of turbulent non-Newtonian power-law fluid flows. Journal of Non-Newtonian Fluid Mechanics 322. https//doi.org/10.1016/j.jnnfm.2023.105136.

\bibitem{4}
Khashi'ie NS, Waini I, Kasim ARM, Zainal NA, Ishak A, Pop I (2022). Magnetohydrodynamic and viscous dissipation effects on radiative heat transfer of non-Newtonian fluid flow past a nonlinearly shrinking sheet: Reiner–Philippoff model. Alexandria Engineering Journal 61(10):7605--7617.

\bibitem{5}
Pan X, Xu J, Zhu Y (2022). Global existence in critical spaces for non Newtonian compressible viscoelastic flows. Journal of Differential Equations 331:162--191.

\bibitem{6}
Akram S, Athar M, Saeed K, Razia A (2023). Influence of an induced magnetic field on double diffusion convection for peristaltic flow of thermally radiative Prandtl nanofluid in non-uniform channel. Tribology International 187. https//10.1016/j.triboint.2023.108719.

\bibitem{7}
Ishtiaq B, Nadeem S (2023). Contraction/dilation flow analysis of non-Newtonian fluid in a deformable channel with exponential permeable walls. Chinese Journal of Physics 83:231--241.

\bibitem{8}
Rahman M, Manzur M, Khan M (2016). Mixed convection heat transfer to  modified second grade fluid in the presence of thermal radiation. Journal of Molecular Liquids 223:217--222.

\bibitem{9}
Hayat T, Nawaz S, Alsaedi A, Rafiq M (2016). Impact of second-order velocity and thermal slips in the mixed convective peristalsis with carbon nanotubes and porous medium. Journal of Molecular Liquids 221:434--442.

\bibitem{10}
Aman S, Khan I, Ismail Z, Salleh MZ (2016). Impacts of gold nanoparticles on MHD mixed convection Poiseuille flow of nanofluid passing through a porous medium in the presence of thermal radiation, thermal diffusion and chemical reaction. Neural Computing and Applications 30:789–-797.

\bibitem{11}
Hayat T, Zahir H, Tanveer A, Alsaedi A (2016). Influences of Hall current and chemical reaction in mixed convective peristaltic flow of Prandtl fluid. Journal of Magnetism and Magnetic Material 407:321--327.

\bibitem{12}
Hayat T, Shafique M, Tanveer A, Alsaedi A (2016). Magnetohydrodynamic effects on peristaltic flow of hyperbolic tangent nanofluid with slip conditions and Joule heating in an inclined channel. International Journal of Heat and Mass Transfer 102:54--63.

\bibitem{13}
Ramesh K, Devakar M (2015). Some analytical solutions for flows of Casson fluid with slip boundary conditions. Ain Shams Engineering Journal 6:967--975.

\bibitem{14}
Choudhari R, Gudekote M, Vaidya H, Prasad KV (2018). Peristaltic Flow of Herschel-Bulkley Fluid in an Elastic Tube with Slip at Porous Walls. Journal of Advanced Research in Fluid Mechanics and Thermal Sciences 52(1):63--75.

\bibitem{15}
Gudekote M, Choudhari M, Vaidya H,  Prasad KV (2020). Impact of Variable Transport Properties and Slip Effects on MHD Jeffrey Fluid Flow Through Channel. Arabian Journal for Science and Engineering 45(1): 417--4281.

\bibitem{16}
Vermolen FJ, Vuik C (2019). Numerical Mathematics and Advanced Applications ENUMATH. European Conference Egmond aan Zee. The Netherlands.

\bibitem{17}
Hayat T, Bibi S, Rafiq M, Alsaedi A, Abbasi FM (2016). Effect of an inclined magnetic field on peristaltic flow of Williamson fluid in an inclined channel with convective conditions. Journal of Magnetism and Magnetic Material 401:733--745.

\bibitem{18}
Hayat T, Asghar S, Tanveer A, Alsaedi A (2017). Outcome of slip features on the peristaltic ﬂow of a Prandtl nanoﬂuid with inclined magnetic field and chemical reaction. European Physical Journal Plus 132(5). https://doi/10.1140/epjp/i2017-11486-8.

\bibitem{19}
Ramesh K (2016). Influence of heat and mass transfer on peristaltic flow of a couple stress
fluid through porous medium in the presence of inclined magnetic field in an inclined asymmetric channel. Journal of Molecular Liquids 219:256--271.

\bibitem{20}
Khan AA, Ellahi R, Gulzar MM, Sheikholeslami M (2014). Effects of heat transfer on peristaltic motion of Oldroyd fluid in the presence of inclined magnetic field. Journal of Magnetism and Magnetic Material 372:97--106.

\bibitem{21}
Hayat T, Abbasi FM, Al-Yami M, Monaquel S (2014). Slip and Joule heating effects in mixed convection peristaltic transport of nanofluid with Soret and Dufour effects. Journal of Molecular Liquids 194:93--99.

\bibitem{22}
Hayat T, Iqbal R, Tanveer A, Alsaedi A (2016). Soret and Dufour effects in MHD peristalsis of pseudoplastic nanofluid with chemical reaction. Journal of Molecular Liquids 220:693--706.

\bibitem{23}
Zhao J, Zheng L, Zhang X, Liu F (2016). Convection heat and mass transfer of fractional MHD Maxwell fluid in a porous medium with Soret and Dufour effects. International Journal of Heat and Mass Transfer 103:203--210.

\bibitem{25}
Alsaedi A, Batool N, Yasmin H, Hayat T (2013). Convective heat transfer analysis on Prandtl fluid model with peristalsis. Applied Bionics and Biomechanics 10:197--208.

\bibitem{26}
Patel M, Surati H, Chanda MS, Timol MG (2013). Models of various non-Newtonian fluids. International e-Journal for Education and Mathematics 2:21--36.

\bibitem{27}
Tzirtzilakis EE (2005). A mathematical model for blood flow in magnetic field. Physics Fluids 17. https//doi:10.1063/1.1978807.

\bibitem{28}
Mao J, Aleksandrova S, Molokov S (2008). Joule heating in magnetohydrodynamic flows in channels with thin conducting walls. International Journal of Heat and Mass Transfer 51:4392--4399.

\bibitem{29}
Kaladhar K, Reddy KM, Srinivasacharya D (2019). Inclined magnetic field, thermal radiation and Hall current effects on mixed convection flow between vertical parallel plates. Journal of Heat Transfer 141. https//doi.:10.1115/1.4044391.

\bibitem{30}
Akbar NS (2014). Blood flow analysis of Prandtl fluid model in tapered stenosed arteries. Ain Shams Engineering Journal 5(4):1267--1275.

\bibitem{31}
Hayat T, Abbasi FM, Ahmad B, Alsaedi A (2014).  Peristaltic Transport of Carreau-Yasuda Fluid in a Curved Channel with Slip Effects. PLoS One 9(4). https//doi:10.1371/journal.pone.0095070.

\bibitem{32}
Ramesh K, Devakar M (2015). The influence of heat transfer on peristaltic transport of MHD second grade fluid through porous medium in a vertical asymmetric channel. Journal of Applied Fluid Mechanics 8(3):351--365.

\bibitem{33}
Mehmood OU, Mustapha N, Shafie S (2012). Heat transfer on peristaltic flow of fourth grade fluid in inclined asymmetric channel with partial slip. Applied Mathematics and Mechanics 33(10):1313–-1328.

\bibitem{34}
Ehsan T, Anjum HJ, Asghar S (2020). Peristaltic flows: A quantitative measure for the size of a bolus. Physica A: Statistical Mechanics and its Applications 553: 124211. https//doi:10.1016/j.physa.2020.124211.

\bibitem{35}
 Eytan O, Elad D (1999). Analysis of Intra-uterine Fluid Motion Induced by Uterine Contractions. Bulletin of Mathematical Biology 61(2):221--238.

\bibitem{38}
 Beavers GS, Joseph DD (1967). Boundary conditions at a naturally permeable wall. Journal of Fluid Mechanics 30(1):197--207.

\bibitem{39}
Crank J (1975). The Mathematics of Diffusion (2nd edition). Oxford University Press.

\bibitem{40}
Ozisik MN (1993). Heat Conduction (2nd edition). Wiley.

\bibitem{41}
Riaz MB, Saddiqa A, Bilal S (2025). Peristaltic transport in Prandtl fluid with diffusion and activation energy aspects by executing numerical simulations. International Communications in Heat and Mass Transfer 163:108742. https://doi.org/10.1016/j.icheatmasstransfer.2025.108742.

\bibitem{42}
Vajravelu K, Sreenadh S, Lakshminarayana P, Sucharitha G (2016). The effect of heat transfer on the nonlinear peristaltic transport of a Jeffrey fluid through a finite vertical porous channel. International Journal of Biomathematics 9(2):1650023. https://doi.org/10.1142/S1793524516500236.

\bibitem{43}
Ajithkumar M, Lakshminarayana P, Vajravelu K (2023). Peristaltic flow of bioconvective Ree–Eyring nanofluid through an inclined elastic channel with partial slip effects. Journal of Applied Physics 134:154701. https://doi.org/10.1063/5.0171422.

\bibitem{44}
Jagadesh V, Sreenadh S, Ajithkumar M, Lakshminarayana P, Sucharitha G (2024). Numerical exploration of the peristaltic flow of MHD Jeffrey nanofluid through a non-uniform porous channel with Arrhenius activation energy. Numerical Heat Transfer, Part A: Applications. https://doi.org/10.1080/10407782.2024.2332477.

\bibitem{45}
Akbar NS, Nadeem S, Lee C (2012). Peristaltic flow of a Prandtl fluid model in an asymmetric channel. International Journal of the Physical Sciences 7(5):687--695.

\bibitem{46}
Ramarao I, Basavaraju PN, Seethappa J (2022). Peristaltic Flow and Heat Transfer Through a Prandtl Fluid in Vertical Annulus. Lecture Notes in Mechanical Engineering. Springer, Singapore. https://doi.org/10.1007/978-981-19-1388-4-16.

\bibitem{47}
Hayat T, Asghar S, Tanveer A (2019). Effects of Hall current and ion-slip on the peristaltic motion of couple stress fluid with thermal deposition. Neural Computing and Applications 31:117--126.

\end{thebibliography}
\end{document}